\documentclass[reprint,aps,prl,longbibliography,superscriptaddress]{revtex4-2}

\usepackage[english]{babel}
\usepackage{dcolumn}

\usepackage{csquotes}
\usepackage{amsmath}
\usepackage{graphicx}
\usepackage[colorlinks=true, allcolors=blue]{hyperref}

\begin{document}

\title{Phylogenetic Corrections and Higher-Order Sequence Statistics in Protein Families: The Potts Model vs MSA Transformer}

\author{Kisan Khatri}
\affiliation{Department of Physics and Center for Biophysics and Computational Biology, Temple University, Philadelphia, PA 19122, USA}
\author{Ronald M. Levy}
\affiliation{ Department of Chemistry and Center for Biophysics and Computational Biology, Temple University, Philadelphia, PA 19122, USA}
\author{Allan Haldane}
\email{allan.haldane@temple.edu}
\affiliation{ Department of Physics and Center for Biophysics and Computational Biology, Temple University, Philadelphia, PA 19122, USA}


\begin{abstract}
Recent generative learning models applied to protein multiple sequence alignment (MSA) datasets include simple and interpretable physics-based Potts covariation models and other machine learning models such as MSA-Transformer (MSA-T). The best models accurately reproduce MSA statistics induced by the biophysical constraints within proteins, raising the question of which functional forms best model the underlying physics. The Potts model is usually specified by an effective potential including pairwise residue-residue interaction terms, but it has been suggested that MSA-T can capture the effects induced by effective potentials which include more than pairwise interactions and implicitly account for phylogenetic structure in the MSA. Here we compare the ability of the Potts model and MSA-T to reconstruct higher-order sequence statistics reflecting complex biological sequence constraints. We find that the model performance depends greatly on the treatment of phylogenetic relationships between the sequences, which can induce non-biophysical mutational covariation in MSAs. When using explicit corrections for phylogenetic dependencies, we find the Potts model outperforms MSA-T in detecting epistatic interactions of biophysical origin.
\end{abstract}

\maketitle

\textit{Introduction - } Machine learning models have made great strides in predicting the functional and structural properties of proteins based on large protein sequence datasets, including the subclass of generative protein sequence models (GPSM) that learn an underlying sequence probability distribution $P(S)$ from a Multiple Sequence Alignment (MSA) to generate new synthetic protein sequences $S$. To function well, GPSMs must capture amino acid patterns in the MSA that implicitly encode information about physical constraints on proteins, enabling the design and detection of hidden protein properties from sequence data\cite{levy2017potts,Bepler2021}. This raises the question of the GPSM functional form that best describes protein biophysics and how to measure this. Here we examine two leading GPSMs in recent focus, the Potts model\cite{weigt2009identification, morcos2011direct, mora2011biological, haldane2016structural, haldane2019influence, levy2017potts, Trinquier2021} and the MSA Transformer (MSA-T)\cite{rao2021msa, Lupo2022}.  We suggest that certain statistical characteristics of individual protein families provide the most reliable metrics to measure GPSM performance if they exclude the biasing effects of phylogenetic relationships between sequences, which do not originate from the fundamental biophysical properties of the protein family.  We find that the Potts model better captures such a statistic, the higher-order MSA statistics which arise due to the potentially large epistatic networks within proteins, when the impact of phylogenetic relationships is carefully accounted for.

\begin{figure}
\centering
\includegraphics[width=1.0\columnwidth]{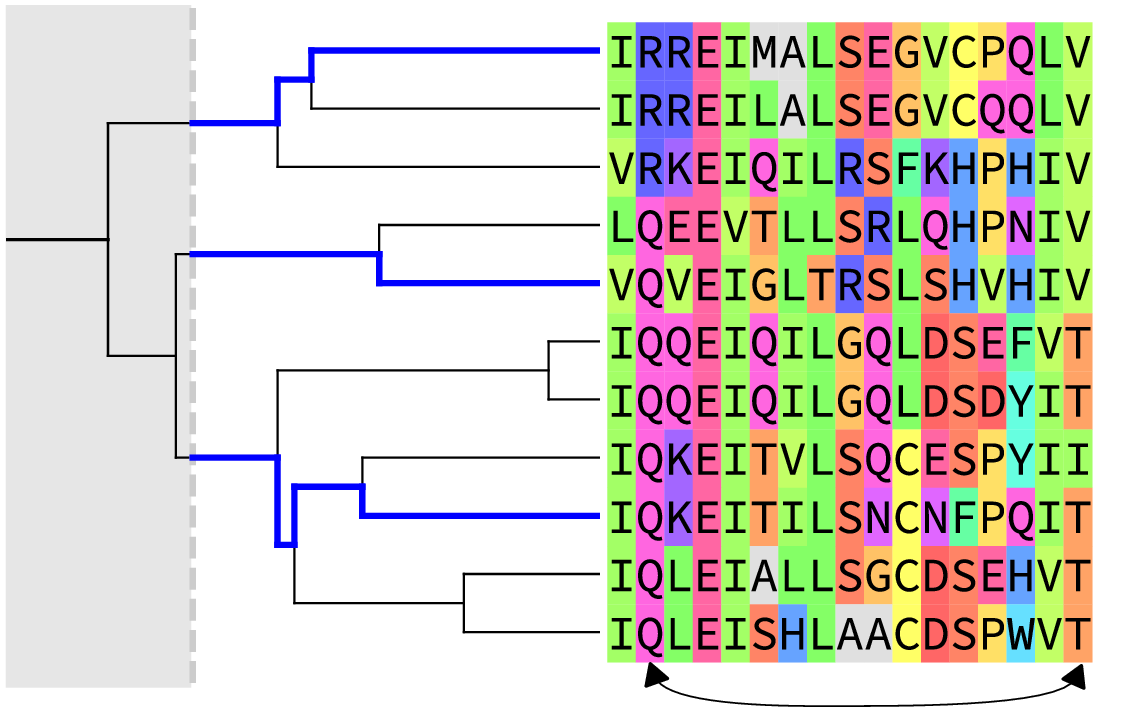} 
\caption{\label{fig:overview} Phylogenetic relationships between sequences in an MSA result in a spurious mutational correlation due to common ancestry, for R/V and Q/T combinations at the illustrated column-pair. Sequence pairs greater than an identity cut-off (diverging to the left of the gray dotted line) approximate \textit{i.i.d.} samples, so that identity filtering to retain 3 sequences labeled in blue gives an unbiased sample showing no correlation with equal frequency of R/V, Q/V and Q/T combinations.}
\end{figure}
\textit{Effect of Phlyogenetic relationships on GPSM evaluation - } Mutational covariation in protein sequences provides a highly informative statistical signal that accurate GPSMs must capture. These arise from multiple factors including: 1) High-fitness sequence motifs and epistasis (mutational cooperativity) underlying biophysical function lead to compensatory mutation pairing patterns between residues. 2) Phylogentic relationships due to recent speciation or gene duplication can distort covariation patterns, causing “excess” counts of patterns from evolutionarily related sequence clusters  as illustrated in Fig. \ref{fig:overview}, and 3) Statistical noise due to the limited number of available distinct natural sequences used during model inference introduces finite-sampling statistical variation in estimated covariation and in GPSM accuracy. The latter two can be considered nuisance factors in uncovering the underlying biophysics driving the first factor of fitness-induced covariation. These covariation signals are measured from MSAs of individual protein families, which consist of homologous proteins with shared function and overall structure.

Despite their quite different architectures and complexity, the Potts model and MSA-T exhibit fundamental commonalities in their design to account for these patterns\cite{martin2024machine}. The Potts model is an interpretable, physics-inspired machine learning model fit to a single protein family, inspired by spin-glasses.  A Potts model for protein family $F$ has  probability distribution $P(S \mid \theta_F) \propto \exp(-\sum_{i<j} J^{ij}_{s_i s_j})$ for sequences $S$ to evolve, with characters $s_i$ at position $i$ ,
and pairwise ``coupling" parameters $\theta_F = \{J^{ij}_{\alpha\beta}\}$ measuring the favorability of having residues $\alpha,\beta$ at positions $i,j$ in a sequence. It models complex higher-order correlations through networks of pairwise interactions. In contrast, MSA-T is a deep learning \cite{lecun2015deep} masked language neural network attention model \cite{vaswani2017attention} trained on MSAs of all available protein families\cite{Rao2021} with about $10^4$ times more parameters, recently used for protein structure and evolution analyses\cite{hawkins2021msa, hong2022s, ma2023retrieved, almalki2024tmhc, cuturello2024enhancing, chen2024learning, chen2024mftrans}. The Potts model has been found to have superior generative accuracy to some other GPSMs including Variational Auto-Encoders and site-independent models \cite{McGee2021}, but a generative method subsequently developed for MSA-T has been reported to better reproduce higher-order statistics \cite{Sgarbossa2023}, though its parameters lack clear physical interpretation and the reasons for this result are unclear. 

GPSMs must distinguish the causes of covariation\cite{de2013emerging}. In the case of the Potts model an identity filtering procedure is critical. The Potts model is trained to reproduce the site-statistics of a training MSA from the protein family, assuming each sequence is an independent and identically distributed (\textit{i.i.d.}) sample. To be \textit{i.i.d}, the sequences can be envisioned to have evolved from a distant ancestral sequence under a common fitness function and mutational process encoded in $P(S \mid \theta_F)$, and enough time must have passed for any statistical correlation with each other to be effectively nil due to mutational saturation. Phylogenetic relationships violate the \textit{independence} assumption, as clusters of orthologs and other closely evolutionarily related sequences are explicitly non-\textit{i.i.d.}. In standard Potts methodology the related sequences are filtered to eliminate phylogenetic redundancy. 

In contrast, MSA-T accounts for phylogenetic relationships through a column-wise attention layer\cite{Lupo2022,chen2024learning}, and is not explicitly trained on MSA statistics but rather on a masked prediction task insensitive to phylogenetic structure, in predicting character states in input MSAs from all protein families at once which were randomly masked. A method has been proposed to use this in a generative fashion for a single protein family by repeatedly masking and resampling an input MSA of that family to produce novel sequences\cite{Sgarbossa2023}.  The MSA-T probability distribution for this generative method is \(P(S\mid \theta, M_F, p) \) for generating sequence $S$, and depends the model's pre-trained parameters $\theta$, an input MSA  $M_F$ for a protein family $F$, and a masking frequency parameter \( p \). 

One goal of this study is to control for how phylogenetic clusters affects GPSM performance. After filtering, the \textit{i.i.d.} MSA will have covariation signals that exclude biases due to clustering and so are presumably biophysical in origin, and should be reproducible in arbitrary subsets of the MSA. This provides a foundation for determining which GPSMs best capture biophysical constraints.

\begin{figure}
\centering
\includegraphics[width=1.0\columnwidth]{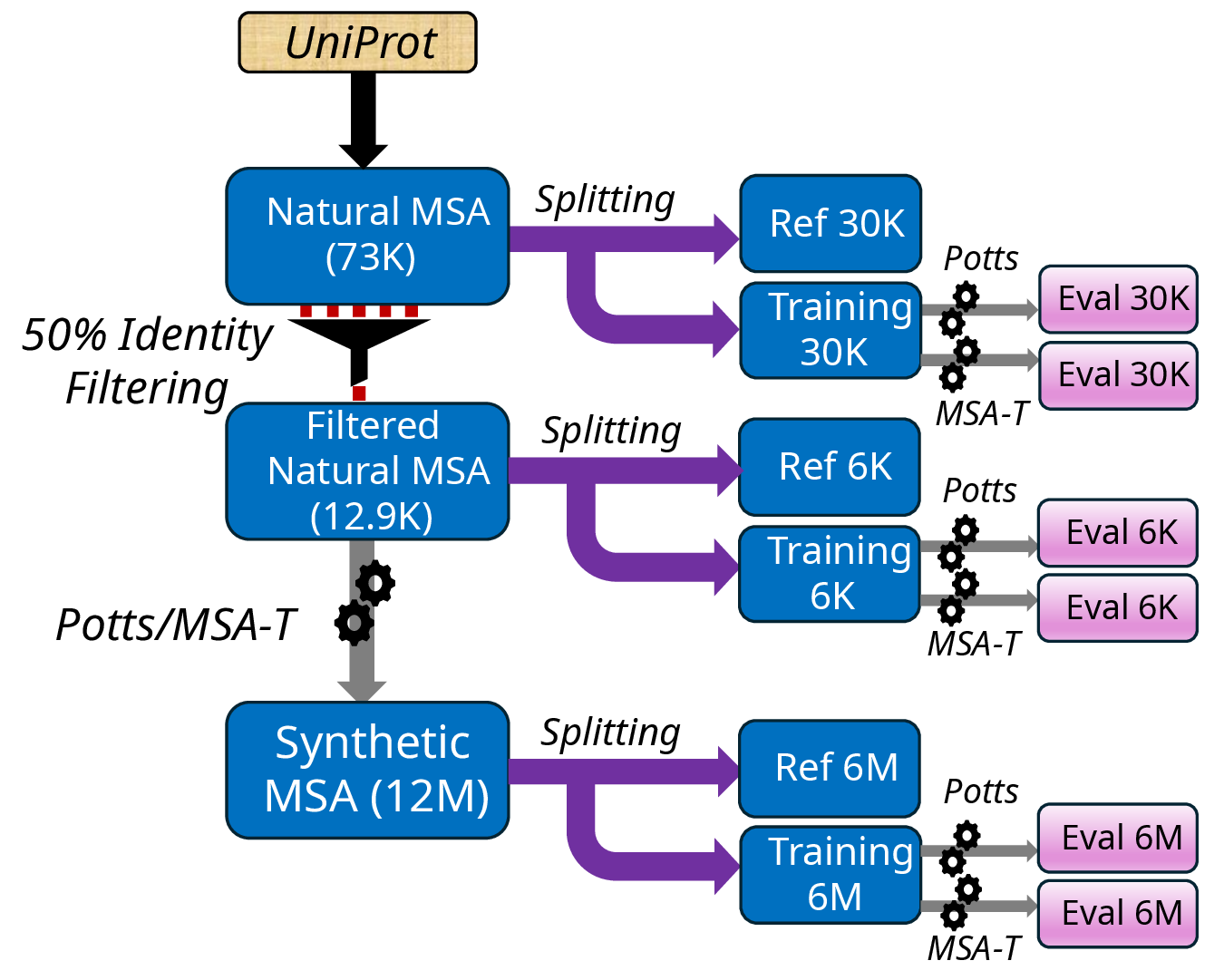} 
\caption{\label{fig:Computational Framework} Overview of statistical tests carried out using the Potts model and MSA-T, to isolate different forms of statistical error. Boxes represent MSAs with different amounts of sequences shown for the RR-domain family, which are then filtered, split, or used to train and generate from the GPSMs (arrows). Our tests measure the statistical difference between the ``evaluation" MSAs and the corresponding ``reference" MSAs.}
\end{figure}
\textit{Statistical methodology -} We perform tests of GPSM performance with and without accounting for phylogeny as outlined in Figure \ref{fig:Computational Framework}. 
 We generate sequences from MSA-T using the default method of Ref. \cite{Sgarbossa2023} using pre-computed parameters $\theta$, and train the Potts model using the high-accuracy Mi3-GPU method\cite{haldane2021mi3}. We also test a very simple GPSM, the Independent model, with distribution $P(S) = \prod_i f^i_{s_i}$ where $f^i_\alpha$ are single-site frequencies for amino-acid $\alpha$ at position $i$ found in the training MSA, which is unable to accurately capture even the pairwise amino-acid frequencies $f^{ij}_{\alpha\beta}$ of the training MSA.  We train the GPSMs on various MSA data described in subsequent sections, generate new synthetic datasets from each GPSM, and then evaluate GPSM performance by comparison of the generated ``evaluation" MSA to a ``reference" MSA using a metric, $r_{20}$, which measures higher-order covariation and provides a more stringent test of GPSM accuracy than pairwise covariation statistics or point-mutant effects\cite{McGee2021}. The training, reference, and evaluation MSAs have the same number of sequences. These tests are designed to isolate the effects of fitness, phylogeny and statistical noise, and to control statistical errors including model specification, out-of-sample, and estimation errors\cite{McGee2021}, such as by splitting each MSA dataset into training and reference MSAs to ensure that no sequences used to train the model are used in its evaluation. We examine two protein families: RR Domain (PF000720), which was previously studied to test MSA-T predictions of higher-order sequence statistics\cite{Sgarbossa2023}, and protein-kinase (PF00069)\cite{McGee2021}. The details of the methodology are provided in the ``End Matter''.

The $r_{20}$ metric measures how well the GPSM predicts the frequency of non-contiguous amino acid ``words" of length $n$ as described in detail previously\cite{McGee2021}. In summary, for each order $n$, 3000 randomly chosen sets of positions of size $n$ are evaluated. For each set of positions, the top 20 most common ``words" at these positions in the reference MSA are found, and the frequency of these words are also computed in both the reference MSA and in the GPSM-generated evaluation MSA. The $r_{20}$ metric is the Pearson correlation between the 20 reference and 20 evaluation frequencies, averaged over the 3000 sets. Focusing on only the top 20 most frequent words restricts the computation to statistically reliable values. This metric measures the ability of the GPSM to model complex chains and networks of interactions within proteins.

To perform identify filtering of MSAs we iteratively find pairs of sequences in the MSA are more than 50\% similar and randomly drop one of the two, because in most protein families distantly related sequences have 10\%-50\% identity, while $>50\%$ suggests recent ancestry and non-\textit{i.i.d.} sequences.  This largely eliminates the influence of phylogenetic clusters on MSA statistics so that most covariation in the training MSA is biophysically induced. For RR-domain, we have 73K sequences, which become filtered to 12.9K sequences, then split into 6K reference and training. For protein-kinase, we have 292K sequence which become filtered to 20K sequences. We also performed our analyses using MSAs filtered at 60\% and 90\% cutoffs to test if the results were sensitive to cutoff, finding qualitatively they were not.
 
\begin{figure}
\centering
\includegraphics[width=1.0\columnwidth]{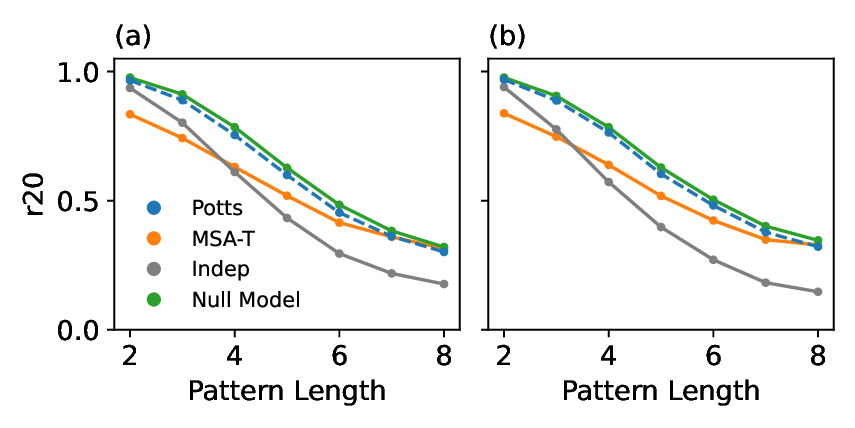} 
\caption{\label{fig: Natural Analysis} ``Natural" GPSM performance test in which the training and reference MSAs are natural sequences filtered by sequence identity to eliminate phylogenetic redundancy, and evaluated using the $r_{20}$ metric for (a) RR-domain (MSAs of 6K) (b) Kinase Protein (MSAs of 10K).}
\end{figure}

\textit{The Potts model outperforms MSA-T after Phylogenetic Pre-processing}- We first tested the performance of the model after filtering the training and reference MSAs, corresponding to the middle section of Figure \ref{fig:Computational Framework}. In Figures \ref{fig: Natural Analysis}(a) (RR domain) and \ref{fig: Natural Analysis}(b) (kinase), we show that the Potts model outperforms MSA-T in this test. 

The $r_{20}$ metric is lower at higher-orders because of greater finite sampling error when measuring the smaller frequencies at these orders, and not because of reduced model accuracy\cite{McGee2021}. The maximum attainable $r_{20}$ metric for a well-specified model subject only to finite-sampling limitations in model evaluation can be found by measuring $r_{20}$ between the training MSA and reference MSA. The Potts model results closely match this null model value, unlike MSA-T. 

\begin{figure}
\centering
\includegraphics[width=1.0\columnwidth]{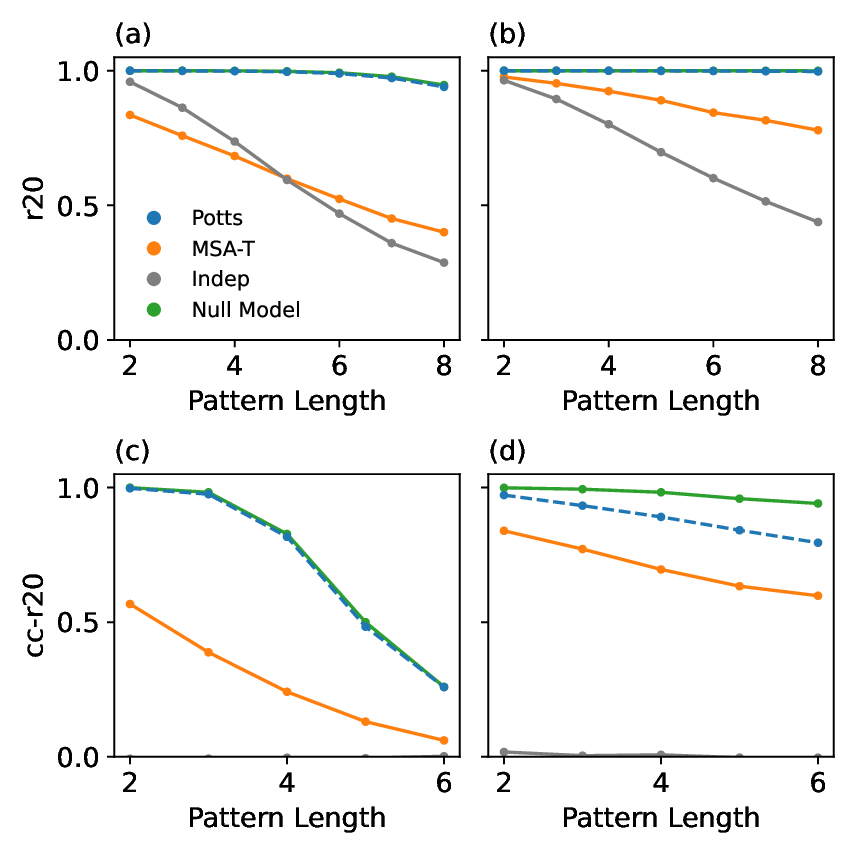} 
\caption{\label{fig:synthetic} ``Synthetic" GPSM performance test for the RR Domain in which large (6M) training and reference MSAs are produced by an initial GPSM, which is a Potts model in (a),(c), and  MSA-T in (b),(d).  The $r_{20}$ metric is used in (a),(b), and a cc-$r_{20}$ metric in  (c) and (d) .}
\end{figure}

The filtered natural MSAs have limited numbers of sequences (12.9K for RR-doamin) which causes significant finite-sampling error at high-orders of marginal. To bypass this limitation we conducted a ``synthetic" test with large training and reference MSAs of 6M sequences. Here, we first trained an initial Potts model on filtered natural MSA and generated two synthetic 6M MSAs from it to serve as reference and training MSAs in the next step. A new Potts, MSA-T, and Independent model were trained on this synthetic \textit{i.i.d.} training dataset. The result shown in Figure \ref{fig:synthetic}a supports that the Potts model outperforms MSA-T when the input sequences are \textit{i.i.d.} even at higher-orders of marginal which are better probed in this test. Interestingly, MSA-T performed worse than a site-independent model for low orders but slightly outperformed it at higher-orders in the case of a synthetic Potts process. This synthetic test might not accurately represent how GPSMs would perform on natural datasets because the reference MSA was generated by the Potts model itself, and the re-fit Potts model should have zero specification error by construction. To address this we performed another test in which the training and reference MSAs were generated by MSA-T from the natural dataset, showing in Figure \ref{fig:synthetic}(b) that the Potts model still outperforms MSA-T. If MSA-T captures biophysical constraints in natural MSAs which the Potts model cannot, we would instead expect lower Potts model performance. Interestingly, MSA-T is unable to reproduce the HOMs when trained on its own generated MSAs, as the $r_{20}$ metric at higher-order are lower than both the null expectation and the Potts result. We investigated this by testing the alternative generation algorithms presented in Ref. \cite{Sgarbossa2023}, as well as changing the acceptance rate $p$ and other variations, but always found qualitatively similar results suggesting a general limitation of MSA-T.

An even more stringent test of the GPSMs is to measure a ``connected correlation $r_{20}$" (cc-$r_{20}$) metric which compares connected correlations, which are higher-order generalizations of the pairwise amino-acid covariation values $C^{ij}_{\alpha\beta} = f^{ij}_{\alpha\beta} - f^i_\alpha f^j_\beta$\cite{McGee2021}. A site-independent model should have zero connected correlation at all orders. In Figures \ref{fig:synthetic}(c) and \ref{fig:synthetic}(d), MSA-T scores significantly worse using cc-$r_{20}$ than the Potts model. Interestingly, in \ref{fig:synthetic}(d), where synthetic data was generated by MSA-T, the Potts model does not match the null expectation, possibly indicating MSA-T introduces statistical patterns beyond pairwise interactions. However, in natural sequence tests (Fig \ref{fig: Natural Analysis}), the Potts model matched the null expectation, suggesting such patterns may not exist in natural datasets. MSA-T also tends to generate less variation than the Potts model, explaining why the null result is larger in Figure \ref{fig:synthetic}(c) compared to \ref{fig:synthetic}(d).

\begin{figure}
\centering
\includegraphics[width=1.0\columnwidth]{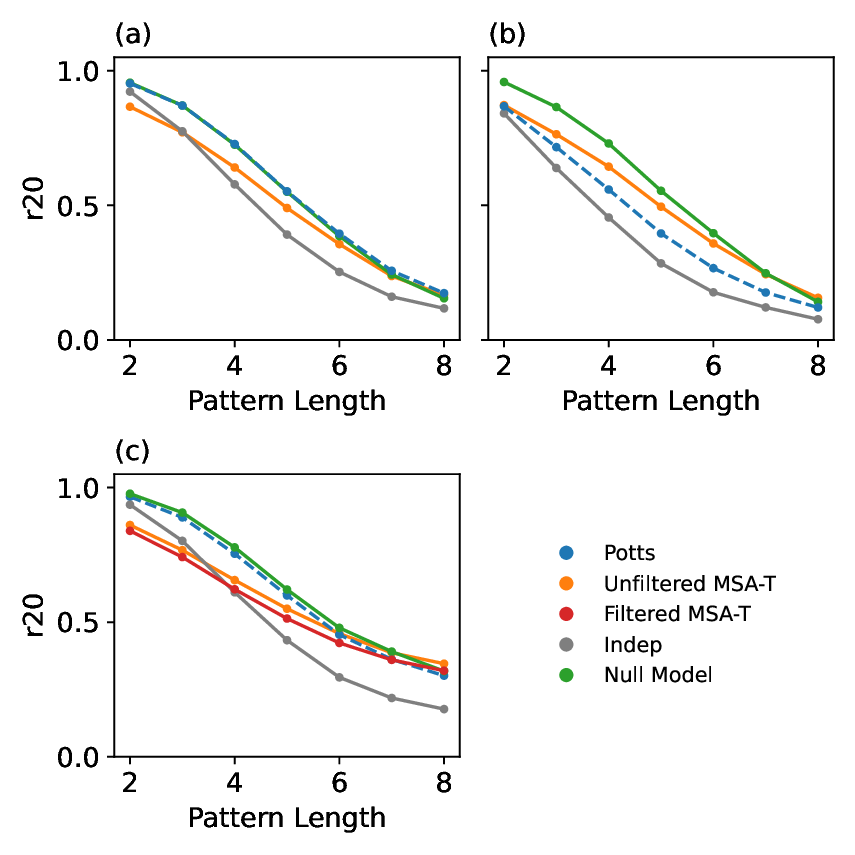} 
\caption{\label{fig:unfiltered} Tests of the impact of identity filtering, for the RR-domain family (a) Models using 30K unfiltered natural sequences for both training and evaluation. (b) The Potts and Independent models trained on 6K filtered natural sequences; MSA-T was trained on 30K unfiltered sequences. The reference MSA is 30K unfiltered. (c) $r_{20}$ value for all models, evaluated against 6K filtered natural sequences.}
\end{figure}

\textit{The Potts model outperforms MSA-T when trained using phylogenetically redundant MSAs - } In \cite{Sgarbossa2023}, it was suggested that MSA-T may have an advantage by being insensitive to phylogenetic structure due to its column-wise attention layers, avoiding the need for identity-filtering which has high computational complexity \(O(N^2)\) for MSA depth $N$. In \cite{Sgarbossa2023}, MSA-T was trained and evaluated on randomly divided MSAs without identity filtering, and the results showed that it outperformed the Potts model, according to $r_{20}$ tested against the unfiltered reference MSA, which we reproduced in Figure \ref{fig:unfiltered}(b). However the Potts model used in that test implicitly performed identity filtering as an internal step. While this can be reasonable if treating the Potts software as a black-box GPSM, it should be expected that a model trained on a filtered MSA will have a lower $r_{20}$ when evaluated with unfiltered reference MSA.  This suggests to test performance when the Potts model is both trained and evaluated on unfiltered MSAs. We find the Potts model outperforms MSA-T in this case (Figure \ref{fig:unfiltered}(a)), suggesting that the Potts model is able to model the component of correlations caused by phylogenetic clusters, to some degree. We expect that such correlations are not biophysical in origin, so that while intriguing this test is inappropriate for testing which GPSM best captures biophysical constraints.

Instead, we suggest that using filtered MSAs for model evaluation will give the best measure a GPSM’s ability to capture the underlying biophysical fitness function, as this will minimize the effects of phylogenetic clustering which introduce sequence correlations driven by non-selective parameters like speciation rates and experimental sampling bias, as discussed above.  In Figure \ref{fig:unfiltered}(c) we compare multiple GPSMs using a filtered reference MSA, and in particular compare MSA-T performance using either filtered and unfiltered training MSAs. MSA-T shows similar performance either way, suggesting it implicitly corrects for phylogenetic structure in the training MSA, if present. However, when the Potts model is trained using a filtered MSA it generally outperforms MSA-T and closely matches the null expectation for a well-specified model. This supports the conclusion that the Potts model more accurately captures features of the underlying biophysical fitness function, as measured by $r_{20}$, than MSA-T.

\textit{Discussion - } The impact of phylogenetic relationships on GPSMs and protein covariation analysis has been recognized since early Potts studies\cite{lapedes1999correlated, Dutheil2012,rao2020transformer,Colavin2022}. Various methods have been proposed to address its confounding effects, such as the “Average Product Correction” (APC)\cite{Dunn2008, Buslje2009} and identity-weighting \cite{weigt2009identification}, or, in profile-HMMs used by HMMER\cite{hmmer} which are a form of GPSMs, by weighting like the Henikoff scheme\cite{henikoff1992amino}. Spectral decomposition of the pairwise covariation matrix \cite{Qin2018}, central to Potts inference, shows its low eigenmodes are influenced by phylogeny while high eigenmodes capture biophysical mutational couplings due to epistasis, and that Potts inference is insensitive to low eigenmodes. This insensitivity has also been found empirically when predicting biophysical ``contacts" in proteins \cite{Horta12021}. This suggests the Potts model may accurately model biophysical interactions even if identity filtering does not completely account for phylogenetic clustering. MSA-T is also designed to account for phylogeny through column attention heads, and it has been found that some attention heads effectively detect sequence relationships\cite{Lupo2022,chen2024learning}.

These results are consistent with previous studies comparing the ability of Potts models and other GPSMs to capture other aspects of protein data including contacts in observed 3d structures of proteins\cite{Bhattacharya2021,hong2022prot}, experimental fitnesses or fitness changes upon mutation\cite{cocco2024functional,hawkins2021msa} which find the architecturally simple Potts model performs favorably. For instance, a previous comparison found that the Potts model outperforms MSA-T for contact prediction if the input data has phylogenetic structure removed\cite{Lupo2022}.

We hypothesize the Potts model outperforms MSA-T in capturing biophysical constraints because: (1) it is trained on a specific protein family while MSA-T is trained on all families; (2) it is directly trained to reproduce pairwise sequence statistics, whereas MSA-T is trained for a masked learning task and so its predictions of marginals are unsupervised; and (3) the Potts model generates sequences with the same diversity as the training MSA, while MSA-T has a free parameter (``replacement rate”) making unclear which value to choose\cite{Sgarbossa2023}. These findings imply the Potts model best captures functional and structural protein constraints despite its architectural simplicity, and highlight the importance of carefully decomposing the origins of covariation, not only when training GPSMs but also during evaluation and in their practical use in understanding the biophysical properties of proteins.
\\

{\textit Acknowledgements - } This research was supported by National Institutes of Health (NIH) grant number R35-GM132090, and  by NIH Computer Equipment Grant (OD020095). Gratitude is also expressed to the OWLSNEST high performance cluster at Temple University for its computing support in this project.

\bibliographystyle{apsrev4-2}
%


\newpage
\clearpage
\section{End Matter: Extended Methods}
\subsection{Natural MSA dataset preparation}  

For the RR domain and protein-kinase families, we constructed MSAs using HHblits\cite{remmert2012hhblits} to search the Uniclust database version $2023\_02$ \cite{uniclust}. For RR Domain, we used the PF00072 seed MSA from Pfam database \cite{el2019pfam}; for Kinase, we used PF00069 with pseudogenes removed as performed in\cite{McGee2021}, resulting in 73,062 raw sequences for RR Domain and 291,731 raw sequences for Kinase.

\subsection{Natural MSA Preprocessing }

We used identity filtering with a 50\% identity threshold, resulting in MSAs of 12,906 sequences for RR domain and 20,000 sequences for kinase. The RR domain was split into 6,000 sequence training and reference MSAs, and the Kinase into 10,000 sequence training and reference MSAs.

When dividing unfiltered natural MSAs into training and reference sets, there can be statistical dependencies that may lead to underestimating out-of-sample error and overestimating $r_{20}$. To address this, we filter the MSA at a 50\% identity threshold before splitting into training and reference halves. We then assign each unfiltered sequence in the original natural MSA to either training or reference set, depending on most similar sequence, ensuring closely related sequences are grouped together. This procedure minimally affects $r_{20}$.

\subsection{Model Training}

For MSA-T, We tested both ``standard" and ``alternative" sequence generation methods from \cite{Sgarbossa2023}, with some modifications.

In the standard method, we divided the input MSA into 600 sequence batches and iteratively performed the masked MSA prediction task for 200 rounds per batch with a masking rate of $p=0.1$ and default options, using the ``greedy" sampling strategy. Sequences with uninitialized characters $(<cls>)$ were discarded, and new sequences were regenerated to consist only of amino acid characters or gaps. Although convergence measures plateaued after 200 rounds, as noted in \cite{Sgarbossa2023}, continued iteration led to an accumulation of ``gap" characters, especially at the sequence ends, which pronounced more at higher $p$ values. We attribute this artifact in MSA-T predictions to missing terminal sequences in the training data, due to shotgun sequencing, bioinformatics misannotation, or evolutionary length variation in protein termini. To address this effect biasing the higher-order correlations, we modified the iterative masking procedure so that gap characters were never masked and never generated at masked positions, preserving the input MSA’s gap characters. We computed bivariate  marginals of input MSA with a pseudocount as described in Ref. \cite{haldane2021mi3} of scale 1/N, where N is the input MSA depth. For training of the Potts model, we used the Mi3-GPU inference software\cite{haldane2021mi3}.

In the ``alternative" generation procedure, each sequence is iterated separately with 599 randomly drawn sequences from the input MSA, using the ``logits" masked sampling strategy. We modified this to preserve gap structure and fix issues where masked positions were treated as unmasked.

\subsection{Higher-Order Marginal (HOM)}

We use the HOM\_r20 package\cite{McGee2021} to calculate the precision of the model in reconstructing the higher-order marginals (HOMs) of MSA. We calculate the $r_{20}$ value \cite{McGee2021} for each HOM of the second to eighth order using 3000 randomly selected column sets of natural and evaluation MSAs. We compute the 20 most frequent amino acid subsequences for each position in training and evaluation MSAs, then calculate the Pearson correlation (r) between these frequencies. The average r, called the $r_{20}$ metric, indicates how well the generated MSA reconstructs the sequence statistics and higher-order mutational patterns of the natural MSA.

\subsection{Synthetic MSA Generation and Analysis} 

We generate ``synthetic" evaluation MSA datasets using models generatively. First, with the protein-kinase Potts model fit to the 12.9K filtered MSA of the RR Domain, we created 6M ``synthetic training" MSAs and two sets of 6M ``synthetic reference" MSAs using Mi3GPU using MCMC\cite{haldane2021mi3}. We then trained new Potts, MSA Transformer, and Independent models on the 6M synthetic training MSA and generated synthetic ``evaluation" MSAs containing 6M sequences for all models. We repeated a similar process for the MSA Transformer using the same input. This approach addresses finite sampling limitations by generating MSAs with any desired number of sequences.

\end{document}